\documentclass[fleqn,usenatbib]{mnras}

\usepackage[T1]{fontenc}

\DeclareRobustCommand{\VAN}[3]{#2}
\let\VANthebibliography\thebibliography
\def\thebibliography{\DeclareRobustCommand{\VAN}[3]{##3}\VANthebibliography}

\usepackage{graphicx}	
\usepackage{amsmath}	
\usepackage{amssymb}	
\usepackage{bm}	

\newcommand{\kpc}{\hbox{$\mathrm{kpc}$}}
\newcommand{\kms}{\hbox{$\mathrm{km \, s^{-1}}$}}
\newcommand{\kmskpc}{\hbox{$\mathrm{km \, s^{-1} \, kpc^{-1}}$}}

\title[Breathing motion along the Local Arm]{Growing Local arm inferred by the breathing motion}

\author[T. Asano et al.]{
Tetsuro Asano,$^{1}$\thanks{E-mail: t.asano@astron.s.u-tokyo.ac.jp}
Daisuke Kawata,$^{2,4}$
Michiko S. Fujii,$^{1}$ and
Junichi Baba$^{3,4}$
\\
$^{1}$Department of Astronomy, Graduate School of Science, 
    The University of Tokyo, 7-3-1 Hongo, Bunkyo-ku, Tokyo, 113-0033, Japan\\
$^{2}$Mullard Space Science Laboratory, University College London,
Holmbury St. Mary, Dorking, Surrey RH5 6NT, UK\\
$^{3}$Amanogawa Galaxy Astronomy Research Center, Kagoshima University, 1--21--35 Korimoto, Kagoshima 890-0065, Japan\\
$^{4}$National Astronomical Observatory of Japan, Mitaka-shi, Tokyo 181-8588, Japan\\
}

\date{Accepted XXX. Received YYY; in original form ZZZ}

\pubyear{2023}

\begin{document}
\label{firstpage}
\pagerange{\pageref{firstpage}--\pageref{lastpage}}
\maketitle

\begin{abstract}
Theoretical models of spiral arms suggest that the spiral arms provoke a vertical bulk motion in disc stars. 
By analysing the breathing motion, a coherent asymmetric vertical motion around the mid-plane of the Milky Way disc, with  \textit{Gaia} DR3, we found that a compressing breathing motion presents along the Local arm.
On the other hand, with an $N$-body simulation of an isolated Milky Way-like disc galaxy, we found that the transient and dynamic spiral arms induce compressing breathing motions when the arms are in the growth phase, while the expanding breathing motion appears in the disruption phase. The observed clear alignment of the compressing breathing motion with the Local arm is similar to what is seen in the growth phase of the simulated spiral arms. Hence, we suggest that the Local arm's compressing breathing motion can be explained by the Local arm being in the growth phase of a transient and dynamic spiral arm.
We also identified the tentative signatures of the expanding breathing motion associated with the Perseus arm and also the Outer arm coinciding with the compressing breathing motion. This may infer that the Perseus and Outer arms are in the disruption and growth phases, respectively. 
\end{abstract}

\begin{keywords}
	Galaxy: disc -- Galaxy: kinematics and dynamics -- Galaxy: structure. 
\end{keywords}


\section{Introduction}
Since \citet{1952AJ.....57....3M} first identified spiral arms in the Milky Way (MW), their geometry, structure, and kinematics have been studied with various tracers \citep{2017AstRv..13..113V, 2020RAA....20..159S}.
Observations of OB stars and H\,II regions suggest that the MW is a four-arm spiral galaxy \citep[e.g.,][]{1976A&A....49...57G, 2014MNRAS.437.1791U}.
The very long baseline interferometry (VLBI) observations can measure trigonometric parallaxes of molecular masers associated with young massive stars and provide accurate distances to them \citep{2014ApJ...783..130R, 2019ApJ...885..131R, 2020PASJ...72...50V}. Based on these observations which support the four-arm model,
the Sun is located between two main arms: the Perseus arm and the Sagittarius-Carina arm, and also close to a minor arm called the Local arm.
The distribution of classical Cepheids shows similar pictures about the spiral geometry \citep{2019Sci...365..478S}.
According to \citet{2014A&A...569A.125H}, both three-arm and four-arm logarithmic spiral models can fit the tracers of H\,II regions, giant molecular clouds and 6.7 GHz methanol masers.
On the other hand, observations of old stars in IR wavelength favour a two-stellar-arm model \citep{2000A&A...358L..13D, 2005ApJ...630L.149B, 2009PASP..121..213C}.
It is still challenging to fully map the global structures of spiral arms in the Galactic disc.

In a few kpc ranges from the Sun, we can study properties of the spiral arms based on the stellar distribution and kinematics thanks to the precise astrometry of \textit{Gaia} \citep{2016A&A...595A...1G}. 
\citet{2019ApJ...882...48M} cross-matched the \textit{Gaia} second data release \citep[DR2;][]{2018A&A...616A...1G} catalogue with the Two Micron All Sky Survey Point Source Catalogue \citep[2MASS PSC;][]{2006AJ....131.1163S} and detected the overdensity of relatively old ($\sim 1$ Gyr) stars at the Local arm. \citet{2021A&A...651A.104P} mapped the overdensity of the young upper main sequence stars using \textit{Gaia} EDR3 \citep{2021A&A...649A...1G} data. They estimated that the length of the Local arm is at least 8 kpc, which indicates that the Local arm is more significant than previously thought. 
Face-on maps of the mean in-plane, i.e. radial and tangential, velocities \citep{2020ApJ...900..186E, 2022MNRAS.512.1574M, 2023A&A...674A..37G, 2023arXiv231006831A} and the median radial action \citep{2023A&A...670L...7P} highlight spiral features.

Spiral arms affect not only planar motion but also vertical motion. They are considered to induce vertical bulk motion called the breathing mode \citep{2014MNRAS.443L...1D, 2014MNRAS.440.2564F, 2016MNRAS.457.2569M, 2016MNRAS.461.3835M, 2022MNRAS.517L..55K, 2022MNRAS.516.1114K}.
The breathing mode is coherent vertical oscillation whose vertical velocity field is antisymmetric about the mid-plane. 
Previous studies \citep[e.g.,][]{2012ApJ...750L..41W, 2013MNRAS.436..101W, 2018A&A...616A..11G, 2019MNRAS.490..797C, 2020MNRAS.491.2104W, 2020A&A...634A..66L,  2022MNRAS.511..784G, 2022A&A...668A..95W, 2023A&A...678A.111A} reported the breathing mode in the Galactic disc. \citet{2022MNRAS.511..784G}  discovered that the observed breathing amplitude increases with the height from the mid-plane. This trend is expected from the spiral-induced breathing mode \citep{2014MNRAS.443L...1D}. 
\citet{2022A&A...668A..95W} reported a tentative evidence of the compressing breathing motion aligned with the Local arm.

The phase spirals are also linked to the bending and breathing modes.
\citet{2018Natur.561..360A} identified a one-arm phase spiral in the \textit{Gaia} DR2 data \citep{2018A&A...616A...1G}, while \citet{2022MNRAS.516L...7H} discovered a two-arm phase spiral in the inner Galaxy in the \textit{Gaia} DR3 data \citep{2023A&A...674A...1G}.
These are associated with the bending and breathing modes, respectively. 
The bending mode can be reasonably explained by the passage of the Sagittarius dwarf galaxy. 
The breathing mode, on the other hand, can be substantially excited by the external perturbation only when the vertical velocity of the perturber is faster than that of the disc stars \citep{2014MNRAS.440.1971W}.
Theoretical approaches \citep{2022ApJ...935..135B, 2023ApJ...952...65B} based on the linear perturbation theory suggest that the bending mode dominates over the breathing mode in the solar neighbourhood and in the inner Galaxy in the case of the Sagittarius dwarf's passage.  The two-arm phase spiral is likely to originate from internal perturbation, possibly induced by spiral arms \citep{2022MNRAS.516L...7H, 2023MNRAS.524.6331L}. 

In this \textit{Letter}, we report that the compressing breathing mode is clearly associated with the Local arm in Section~\ref{sec:gaia}. In Section~\ref{sec:n_body}, we compare the observed breathing pattern with that seen in an $N$-body model of MW-like disc galaxy. We demonstrate that the breathing motion can infer the evolution phases of the spiral arms, and the compressing motion observed in the Local arm can be interpreted as the Local arm being in the growth phase. We provide our conclusion 
in Section~\ref{sec:conclusion}.

\section{The breathing motion in the Milky Way disc}\label{sec:gaia}
\subsection{The \textit{Gaia} data}
From the \textit{Gaia} DR3 \citep{2023A&A...674A...1G} catalogue, we select the stars that satisfy the following criteria: (1) the relative error in parallax is less than 20\%, (2) the renormalised unit weight error (\texttt{ruwe}) is less than 1.4, (3) the radial velocity is not null.
We use $1/\varpi$ as a distance to a star from the Sun, where $\varpi$ is the zero-point-corrected parallax \citep{2021A&A...649A...4L}. 
Here we exclude the data lacking the parameters required for the correction, such as the effective wavenumber (\texttt{nu\_eff\_used\_in\_astrometry}). 
Our resulting sample comprises 26,218,611 stars.
We convert the heliocentric coordinate to the Galactocentric coordinate using \texttt{astropy.coordinates} from \textsc{astropy} Python package \citep{2022ApJ...935..167A}. Here we assume that the Sun is located at $(x, y, z)=(-8.277, 0, 0.0208)\, \kpc$ \citep{2022A&A...657L..12G, 2019MNRAS.482.1417B} and that its velocity is $(11.1, 251.5, 8.59) \,\kms$ \citep{2010MNRAS.403.1829S, 2020ApJ...892...39R} in the Galactocentric coordinate.

\subsection{Face-on breathing map}
\begin{figure*}
	\begin{center}
		\includegraphics[width=0.4\linewidth]{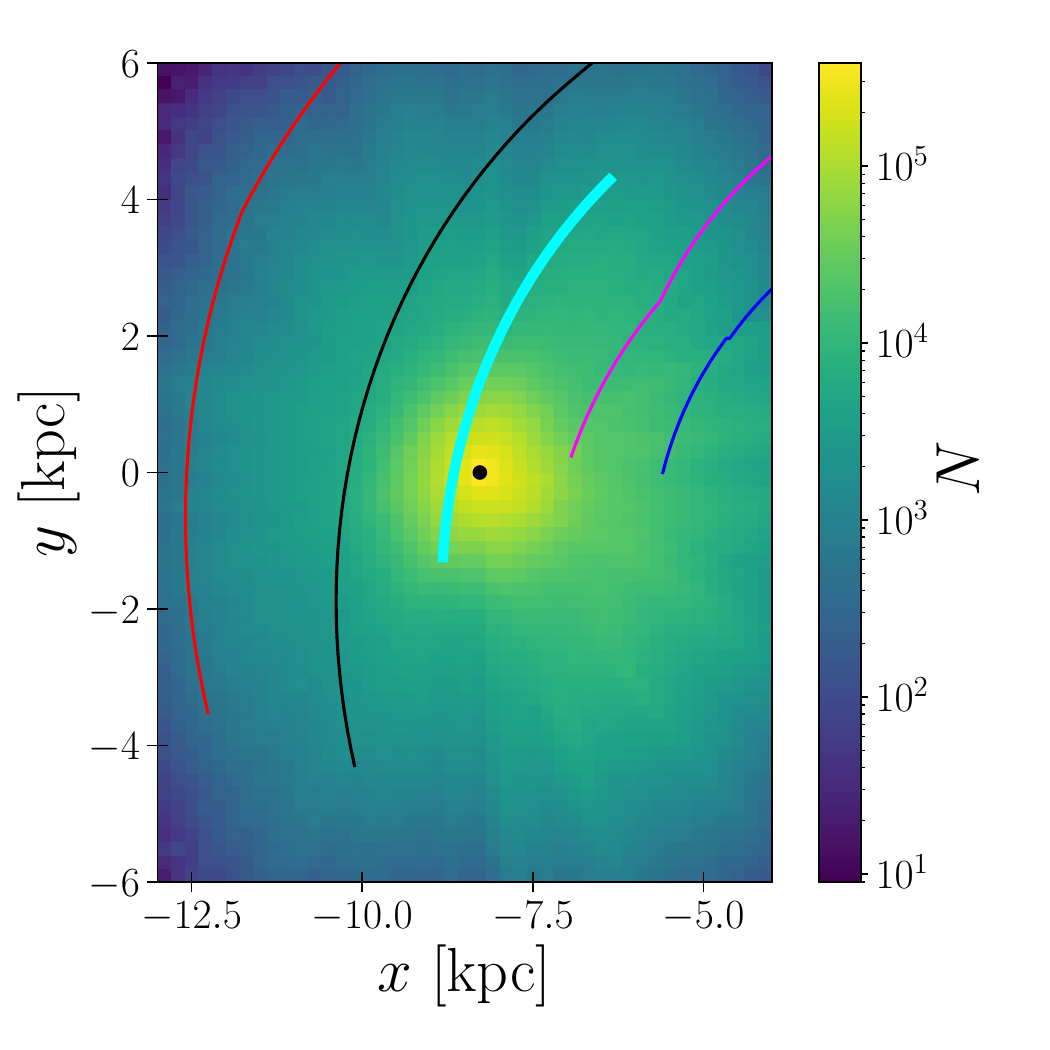}
		\includegraphics[width=0.4\linewidth]{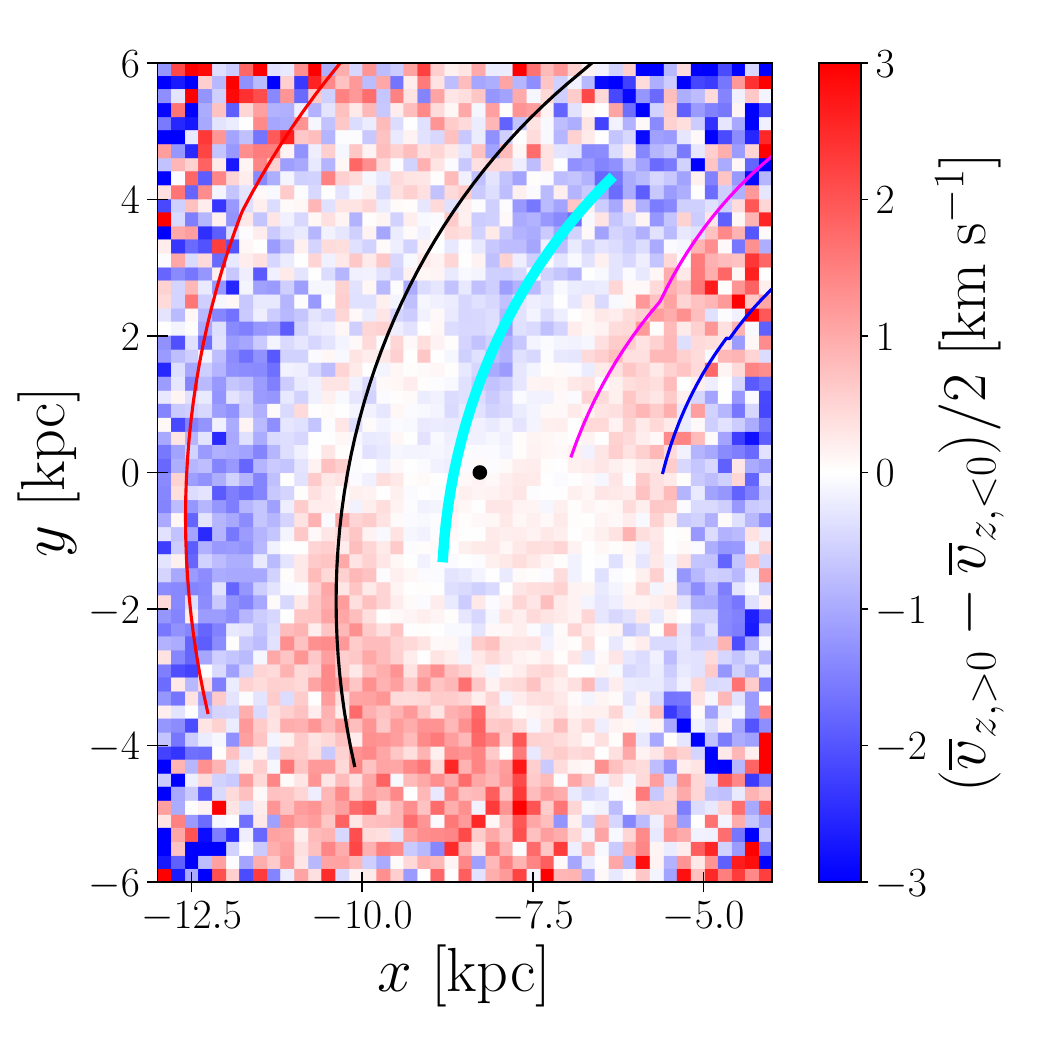}
	\end{center}
	\caption{Face-on maps from the \textit{Gaia} data. \textit{Left:} Number counts. \textit{Right:} Breathing velocity. Solid lines indicate the position of the spiral arms: Outer arm (red), Perseus arm (black), Local arm (cyan), Sagittarius–Carina arm (purple) and Scutum–Centaurus arm (blue), from \citet{2019ApJ...885..131R}. The black dot indicates the position of the Sun.}\label{fig:face_on_Gaia}
\end{figure*}

Fig.~\ref{fig:face_on_Gaia} shows face-on maps of the Galactic disc made from the \textit{Gaia} data. The left panel shows the number counts of the stars in each bin, whose size is $0.2\times 0.2\,\kpc^2$.
The right panel shows the breathing velocity $\frac{1}{2}(\overline{v}_{z,>0}-\overline{v}_{z,<0})$, where $\overline{v}_{z,>0}$ and $\overline{v}_{z,<0}$ are the mean vertical velocities of the stars within $0< z <1.5$ kpc and $-1.5 < z < 0$ kpc, respectively.
The locations of the spiral arms are plotted as follows: Outer arm (red), Perseus arm (black), Local arm (cyan), Sagittarius–Carina arm (purple) and Scutum–Centaurus arm (blue). The locations are based on the spiral arms traced by the massive star-forming regions \citep{2019ApJ...885..131R}, but we add an offset of 0.127 kpc in the radial direction because they assume a Sun-Galactic centre distance different from ours. 
The number counts are significantly affected by the selection bias, and therefore, it is difficult to find intrinsic overdensity or underdensity from this map. 
On the other hand, the right panel exhibits some breathing patterns.
The most distinct structure is a compression area (indicated by the blue colour) extending from the upper right corner of the panel to the point near the Sun. 
Remarkably, this pattern closely aligns with the Local arm highlighted with the cyan line in Fig.~\ref{fig:face_on_Gaia}. This is likely indicating that the Local arm induces the compressing breathing mode.

The identical structure can be seen in the RGB star sample in Fig.~23 of \citet{2023A&A...674A..37G}. They noted that the expanding breathing motion observed in their OB star sample possibly aligns with the Local arm. However, the amplitude of this expansion is smaller than that of the compressing breathing motion seen in the RGB star sample. Additionally, the association with the arm is less distinct in the OB star sample compared to the RGB star sample.
Recently, \citet{2023A&A...678A.111A} reported a breathing mode in the $60^{\circ} \le l \le 75^{\circ}$ direction for the A star sample. They probably have captured the same structure in a different view, but they did not discuss the association with the spiral arm. \citet{2022A&A...668A..95W} also discussed a tentative alignment between the Local arm and the compressing breathing mode. 
They highlighted the offset of the compressing breathing mode from their derived stellar density excess, which is not in the same location as where suggested in \citet{2014ApJ...783..130R}. The true location of the Local arm is still in debate. Because our result shown in Fig.~\ref{fig:face_on_Gaia} presents even more clear alignment with the Local arm position suggested by \citet{2019ApJ...885..131R} than previously shown, in this {\it Letter} we consider that the location of the Local arm is where suggested in \citet{2019ApJ...885..131R}. In Section~\ref{sec:n_body}, we show that with this assumption the Local arm is well explained with the dynamic spiral arm scenario seen in $N$-body simulations.

While the Local arm exhibits a clear connection between the compressing breathing motion, the other spiral arms are more weakly associated with the breathing motion.  The area between the Sagittarius–Carina arm and the Scutum–Centaurus arm exhibits the expanding breathing motion (indicated by the red colour). Outside of the Local arm, a large area of the expansion breathing mode exists, and it looks to partially coincide with the Perseus arm. 
We see compression zones in the outer Galaxy ($R\gtrsim12$ kpc) and in the inner Galaxy ($R\lesssim 5$ kpc).
The Outer arm passes through the outer one. The inner one is not associated with any spiral arms but is near the bar end. One possibility is that it is driven by the bar. Theoretical studies show that both bars and spiral arms can excite breathing mode \citep{2015MNRAS.452..747M, 2016MNRAS.461.3835M, 2023ApJ...952...65B}. However, this is beyond the scope of this {\it Letter}, and the further studies are encouraged. 

\section{Breathing motion in $N$-body simulation}\label{sec:n_body}
\subsection{$N$-body simulation}
We use an isolated MW-like $N$-body model, MWaB, simulated by \citet{2019MNRAS.482.1983F}. It consists of a DM halo, a classical bulge, and an exponential disc. They made the initial condition with \textsc{galactics} \citep{1995MNRAS.277.1341K, 2005ApJ...631..838W, 2008ApJ...679.1239W}. 5.1 billion equal-mass particles of $\sim178M_{\sun}$ represent the model.
The halo initially follows the Navarro-Frenk-White (NFW) profile \citep{1997ApJ...490..493N}, whose mass within $\sim200$ kpc and scale radius are $8.68\times10^{11}M_{\sun}$ and 10 kpc, respectively.  
The bulge follows the Hernquist profile \citep{1990ApJ...356..359H}, whose mass and scale radius are $5.43\times10^{9}M_{\sun}$ and 0.75 kpc, respectively.
The density profile of the disc is radially exponential and vertically isothermal (i.e., $\mathrm{sech}^2$ profile). The disc mass, scale radius and scale height are $3.75\times10^{10}M_{\sun}$, 2.3 kpc and 0.2 kpc, respectively
The simulation was performed with the parallel GPU tree-code \textsc{bonsai} \citep{2012JCoPh.231.2825B, 2014hpcn.conf...54B}  for 10 Gyr.
For more details on the simulations, see \citet{2019MNRAS.482.1983F}.

\begin{figure*}
	\begin{center}
		\includegraphics[width=0.33\linewidth]{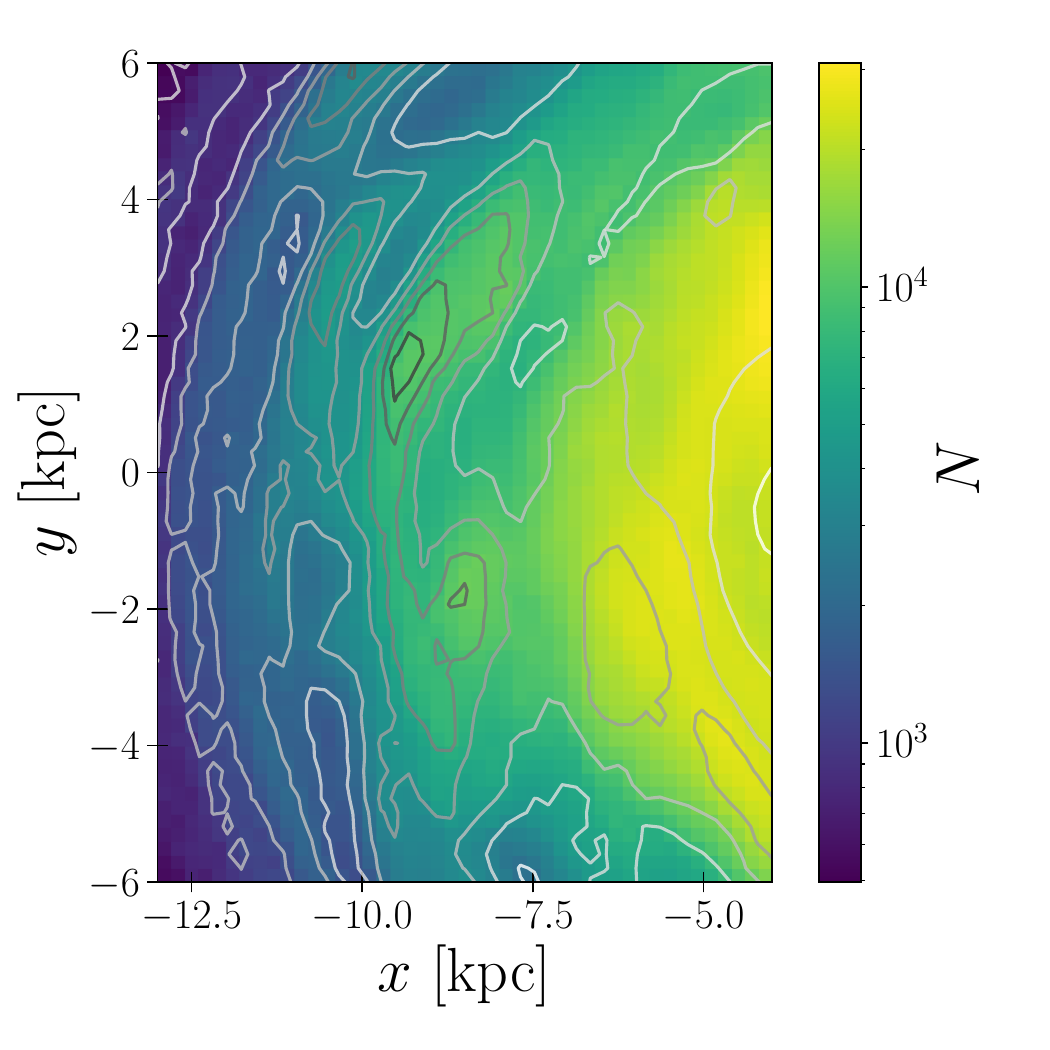}
		\includegraphics[width=0.33\linewidth]{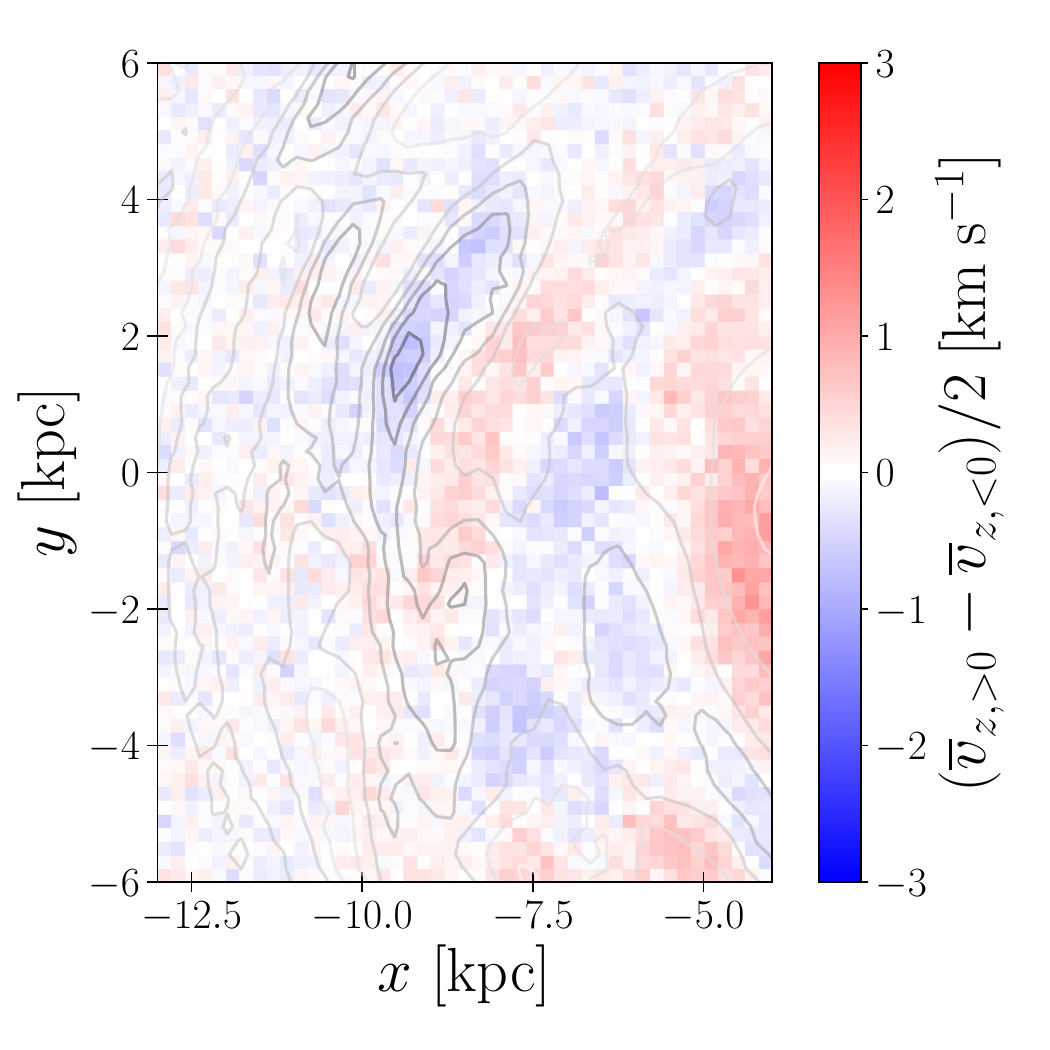}
		\includegraphics[width=0.33\linewidth]{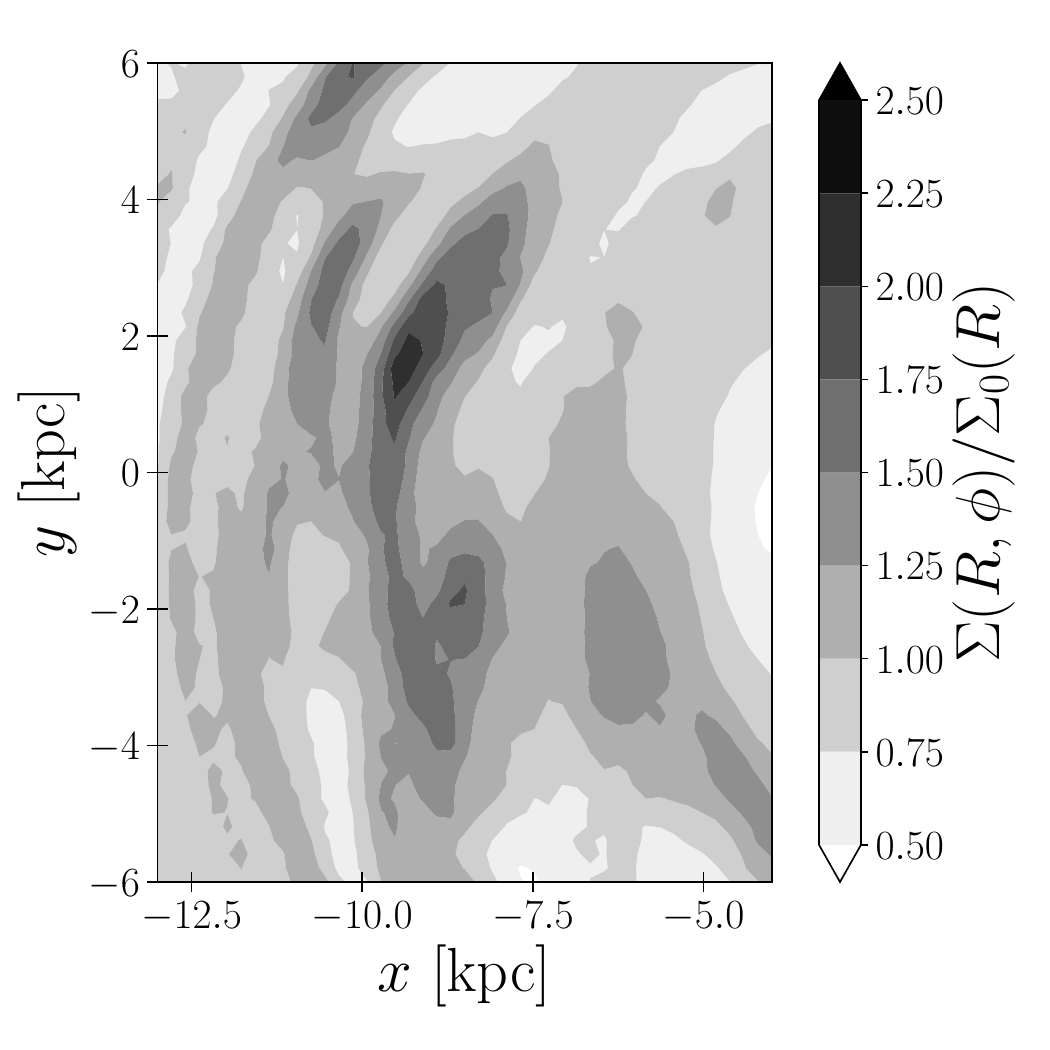}
	\end{center}
	\caption{Face-on maps from the $N$-body model. \textit{Left panel} shows the number counts. \textit{Middle panel} shows the breathing velocity. \textit{Right panel}  shows the contour map of the normalised surface density $\Sigma (R,\phi)/\Sigma_0(R)$, where $\Sigma(R,\phi)$ and $\Sigma_0(R)$ are the 2D surface density and the exponential surface density, respectively. The same contours are shown in the other two panels.}\label{fig:face_on_MWa}
\end{figure*}

\subsection{Breathing motion in $N$-body simulation}
The left and middle panels of Fig.~\ref{fig:face_on_MWa} show a comparable plot to Fig.~\ref{fig:face_on_Gaia} created from the $N$-body model at $t=4.216$ Gyr. The simulated galaxy has a bar, and the angle between its major axis and the $x$-axis is $25^{\circ}$, like the MW. However, the bar is small and does not appear in this plot. 
	Regarding the local stellar kinematics in the solar neighbourhood \citep{2020MNRAS.499.2416A, 2022MNRAS.514..460A} and the global dynamical structure \citep{2019MNRAS.482.1983F}, the snapshots in the later epoch ($t\gtrsim8$ Gyr) of the simulation better reproduce the observed properties such as the local and bulge velocity dispersions of the stars in the MW. However, in this epoch, the spiral arms are fainter due to the dynamical heating of the disc, and the strength of the spiral arms is underestimated. In this \textit{Letter}, we focus on the spiral-induced breathing motion, therefore, we have analysed the earlier snapshots which show prominent spiral structure.
We have selected the snapshot at $t=4.216$ Gyr because it exhibits the spiral arm geometrically most similar to that around the Sun. 
The contour map in the right panel displays the normalised surface density $\Sigma(R,\phi)/\Sigma_0(R)$, where $\Sigma(R,\phi)$ and $\Sigma_0(R)$ are the surface density and the azimuthally averaged surface density profile, respectively. To estimate $\Sigma_0(R)$, we calculate the averaged surface densities for 40 annuli between $R=5$ kpc and 13 kpc and fit them with an exponential profile. The fitted scale radius of 2.4 kpc is comparable to that for the initial condition of the stellar disc. The same contour is overplotted on the left and middle panels.
The middle panel shows the breathing velocity.
The breathing mode is clearly correlated with the spiral arms. The compression mode and the expansion mode appear in the arm and inter-arm regions, respectively. In the outer galaxy, the arms are weak and the breathing velocity is almost zero.

The middle panel of Fig.~\ref{fig:face_on_MWa} shows that the spiral arm seen in the middle of the panel is qualitatively similar to what is seen around the Local arm in the right panel of Fig.~\ref{fig:face_on_Gaia}, although the amplitude of the breathing velocity is smaller than the observed one. 
The \textit{Gaia}'s selection bias can explain this difference.
The intrinsic amplitude of the spiral-induced breathing mode should increase with height from the mid-plane \citep{2014MNRAS.443L...1D}, and this trend has been partially confirmed in the \textit{Gaia} DR2 data \citep{2018A&A...616A..11G, 2022MNRAS.511..784G}. 
In the \textit{Gaia} data, stars at higher heights from the mid-plane are preferentially sampled at the region of the further distance from the Sun, 
because the completeness of the \textit{Gaia} data is lower at low Galactic latitudes at a farther distance due to the dust extinction.
As a result, a higher amplitude of the breathing velocity is expected. 
In fact, the breathing amplitude in the solar neighbourhood ($\lesssim 1$ kpc) is smaller than in the distant region because the data are more complete in the mid-plane. 
In Appendix~\ref{sec:selection_effect}, we quantitatively assess the impact of the selection bias by applying Gaia's selection function to the $N$-body simulation data. As a result, our analysis indicates a tendency for the breathing amplitude to increase due to the influence of Gaia's selection function.

\begin{figure*}
	\begin{center}
		\includegraphics[width=\linewidth]{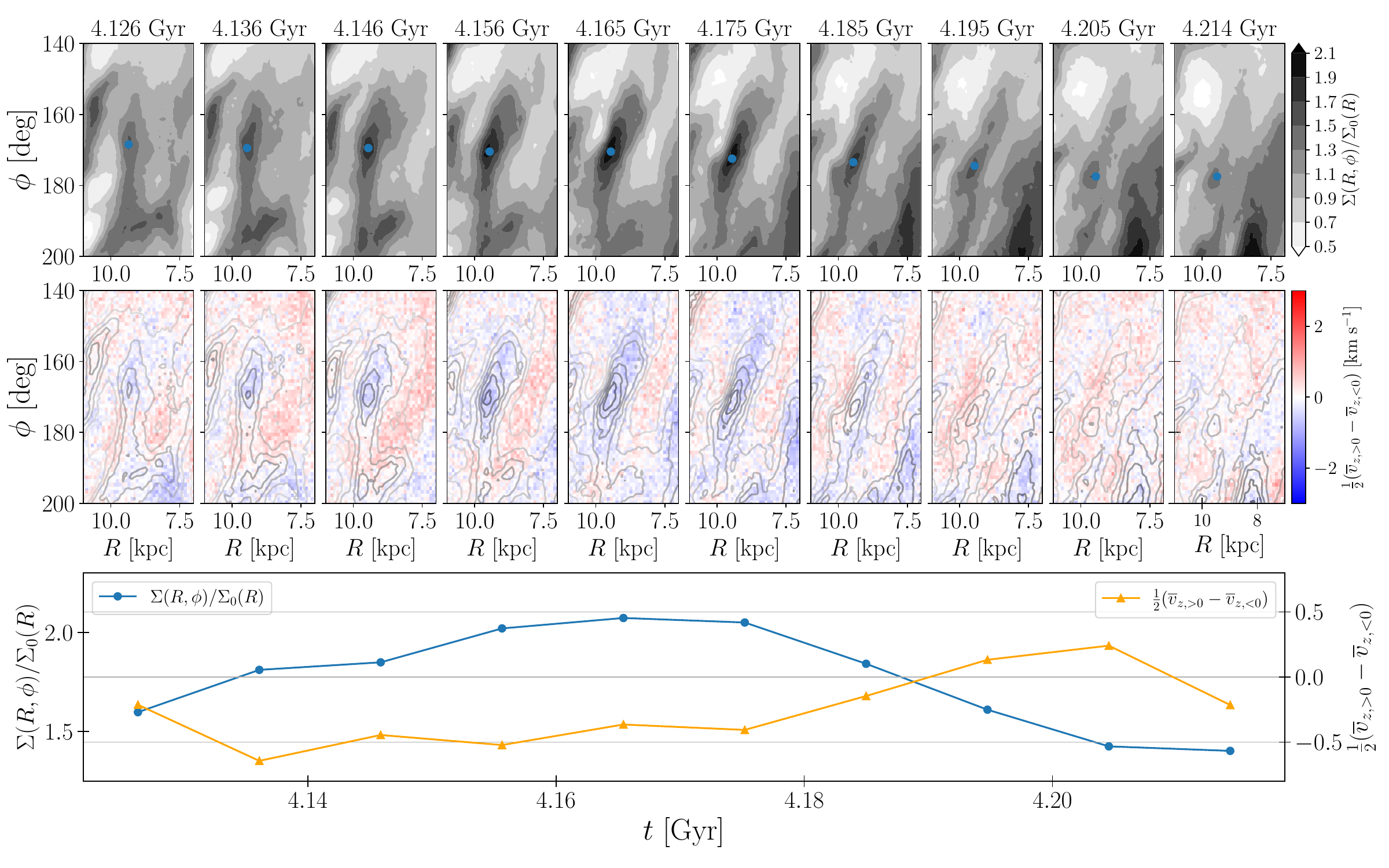}
	\end{center}
	\caption{Time evolution of a spiral arm and the associated breathing motion. \textit{Top panels} shows the normalised surface density in the ten snapshots between $t=4.126$ Gyr and 4.214 Gyr. The density maps are shown in the frame rotating with the circular frequency at $R\sim9$ kpc on $R$-$\phi$ plane. The galaxy rotates in the direction in which $\phi$ decreases. \textit{Middle panels} shows the breathing velocity. The contours of the upper panels are overplotted. \textit{Bottom panel} shows the normalised surface density and the breathing velocity at the points indicated by blue dots in the top panels as functions of time.}\label{fig:time_evolution}
\end{figure*}

\subsection{Time evolution of spiral arms and breathing mode}
We further investigate how the dynamic spiral arms and breathing mode in the $N$-body model evolve with time.
In Fig.~\ref{fig:time_evolution} we present the normalised surface density $\Sigma(R,\phi)/\Sigma_0(R)$ (first row) and the breathing velocity $\frac{1}{2}(\overline{v}_{z,>0}-\overline{v}_{z,<0})$ (second row) around the spiral arm located at the centre of Fig.~\ref{fig:face_on_MWa} for ten sequential time steps ranging from $t=4.126$ Gyr to 4.214 Gyr.  These maps are shown in a frame rotating at an angular velocity of $27\,\kmskpc$, which is the circular frequency at $R\sim9\,\kpc$.  The galaxy rotates in the direction in which $\phi$ decreases. The first (left-hand) and last (right-hand) five columns correspond to the growth and disruption phases of this highlighted spiral arm, respectively.
We trace the peak surface density (indicated by blue points in the top panels) and the breathing velocity there and present their time evolution in the bottom panel of Fig.~\ref{fig:time_evolution}.
The amplitude of the compressing breathing mode at the arm region increases as the arm grows. At the same time, a large expansion of breathing velocity appears at the trailing side of the arm.
The compressing amplitude decreases rapidly during the disruption phase. In the last three snapshots ($t=4.195$, 4.205 and 4.214 Gyr), both the spiral arm and breathing mode become weaker. The breathing mode reaches almost zero or even weakly positive value, i.e., expansion. In the last snapshot, the spiral arm almost disappears, but still identifiable. Hence, the dynamic spiral can also coincide with the expanding breathing mode in the disrupting phase. 

The previous theoretical studies of the breathing mode due to the spiral arms were based on the density-wave theory of the spiral arm. 
\citet{2014MNRAS.443L...1D} showed that stars move towards (away from) the mid-plane when they enter (leave) the spiral arms, if the spiral arms are like density-wave, and the stars rotate with significantly different speed from the pattern speed of the spiral arms. Therefore, the compressing (expanding) breathing mode appears at the trailing (leading) side of the arm inside the corotation radius, and the opposite trend appears outside the corotation radius. They confirmed this phenomenon through investigation in an $N$-body disc model which has grand-designed spiral arms with a well-defined single pattern speed. In an analytical model \citep{2016MNRAS.457.2569M} and test particle simulations \citep{2014MNRAS.440.2564F, 2016MNRAS.457.2569M}, it is also shown that stellar motion is compressing and expanding at the trailing and leading sides, respectively, when the corotation radius is at outside of the disc.

Contrary to these previous studies, in our model, the compressing motion is confined to the arm, and there is no systematic displacement. 
This is because in our simulation the spiral arms are transient and dynamic spiral arms, as seen in many $N$-body simulations \citep[e.g.,][]{1984ApJ...282...61S, 2009ApJ...706..471B, 2013ApJ...763...46B, 2011ApJ...730..109F, 2011ApJ...735....1W, 2012MNRAS.421.1529G, 2012MNRAS.426..167G, 2013ApJ...766...34D, 2015MNRAS.454.2954B}. In the dynamic spiral arms, the spiral-arm features co-rotate with the disc stars at every radius. Hence, as a result, the compression due to the spiral arms coincides with the density contrast of the spiral arm, unlike the offset expected from the density wave-like spiral arms. 

The alignment between the Local arm and the compression mode in the Gaia data in Fig.~\ref{fig:face_on_Gaia} is striking. Based on the result of our simulation and similarity to the observed trend, we are tempted to suggest that the Local arm is a spiral arm strong enough to induce the compressing breathing mode, and it is a co-rotating dynamic arm like what we can find in the $N$-body simulations. 
Then, the compressing breathing motion in the Local arm indicates that the Local arm is in the growth phase of the dynamic arm. 

\section{Discussion and Conclusion}\label{sec:conclusion}
In this \textit{Letter}, we have reported the detection of the breathing mode associated with the spiral arms, especially the striking coincidence of the compression mode in the Local arm, in the \textit{Gaia} data. We have also identified a similar compressing breathing pattern in growing spiral arms in the isolated $N$-body disc model.
If the real spiral arms of the MW are similar to this dynamic nature, we can conclude that the observed compressing breathing motion aligned well with the Local arm infers the growth phase of dynamic spiral arms. The strong compressing breathing feature in the Local arm also indicates that the Local arm is not a minor spiral arm but a major strong arm that influences even the vertical velocity field of the stars.

In contrast to the Local arm, the Perseus arm 
coincides with the expanding breathing mode (see the right panel of Fig.~\ref{fig:face_on_Gaia}).  This implies that the Perseus arm is in the disrupting phase. 
\citet{2018ApJ...853L..23B} discovered that velocities of classical Cepheids around the Perseus arm show a sign of the disruption phase expected from the dynamic spiral arm model \citep{2013ApJ...763...46B, 2015MNRAS.454.2954B}.
The indication from the expanding breathing mode is consistent with this picture. Applying the same inference from our $N$-body simulation result, the Outer arm can be considered as a growth phase, and the Sagittarius-Carina arm and the Scutum-Centaurus arm may be disrupting. 
This may indicate that the MW harbours the different phases of the dynamic spiral arms, which is naturally seen in $N$-body simulations (Funakoshi et al.\ in prep.).
We note that inferring the evolutionary phases of the arms in the inner Galaxy is more challenging, because they are near the bar end, and the combined effect of the spiral arms and the bar makes the vertical stellar motion more complex \citep{2016MNRAS.461.3835M}.
Further studies of the impacts of interference between the bar and spiral arms to the vertical motions are encouraged to comprehensively understand the fascinating nature of the Galactic disc structure.


\section*{Acknowledgements}
We thank the anonymous referee for the useful comments.
This work has made use of data from the European Space Agency (ESA) mission \textit{Gaia} (\url{https://www.cosmos.esa.int/gaia}), processed by the \textit{Gaia} Data Processing and Analysis Consortium (DPAC,
\url{https://www.cosmos.esa.int/web/gaia/dpac/consortium}). Funding for the DPAC has been provided by national institutions, in particular, the institutions participating in the {\it Gaia} Multilateral Agreement. 
Simulations were performed using GPU clusters, HA-PACS at Tsukuba University, Piz Daint at the Swiss National Supercomputing Centre, and Little Green Machine II.
T.A.\ is supported by JSPS Research Fellowships for Young Scientists. 
This work was supported by Grant-in-Aid for JSPS Fellows Number JP22J11943, the UK's Science \& Technology Facilities Council (STFC grant ST/S000216/1, ST/W001136/1) and MWGaiaDN, a Horizon Europe Marie Sk\l{}odowska-Curie Actions Doctoral Network funded under grant agreement no. 101072454 and also funded by UK Research and Innovation (EP/X031756/1).
\section*{Data Availability}
The simulation snapshots are available at \url{http://galaxies.astron.s.u-tokyo.ac.jp}.
The other data and code used in this paper will be shared on reasonable request to the corresponding author.



\bibliographystyle{mnras}
\bibliography{export-bibtex} 




\appendix
\section{Selection effects}\label{sec:selection_effect}
\begin{figure}
	\begin{center}
		\includegraphics[width=0.8\columnwidth]{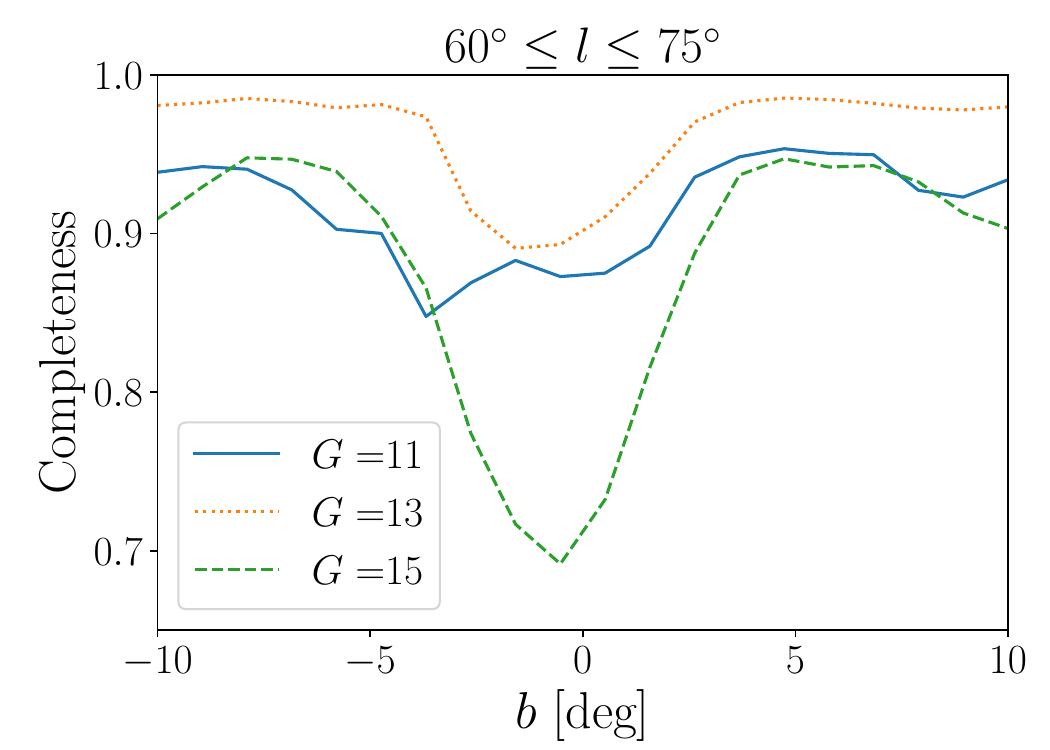}
		\caption{Selection function for the \textit{Gaia} DR3 radial velocity sample at $60^{\circ} \le l \le 75^{\circ}$.}
		\label{fig:selection}
	\end{center}
\end{figure}
\begin{figure}
	\begin{center}
		\includegraphics[width=0.49\columnwidth]{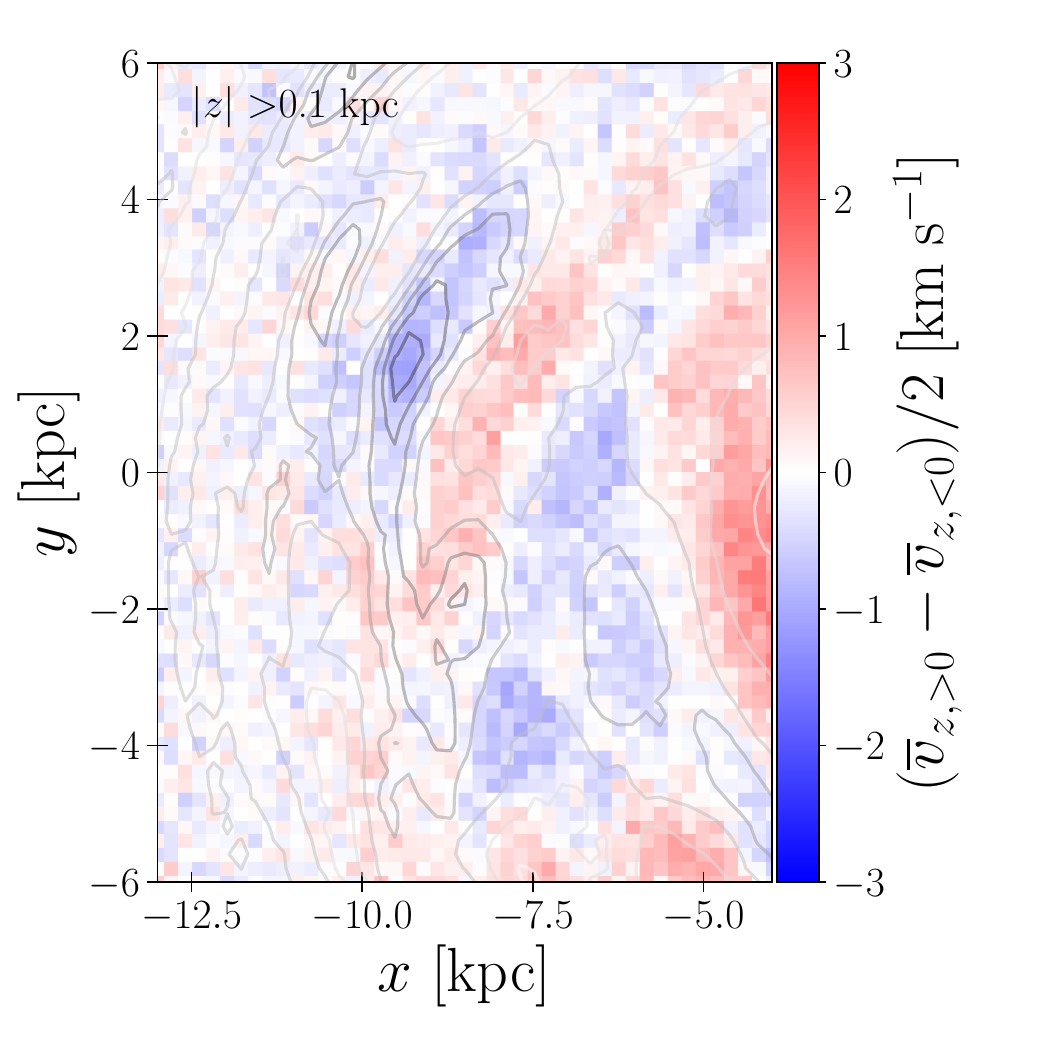}
		\includegraphics[width=0.49\columnwidth]{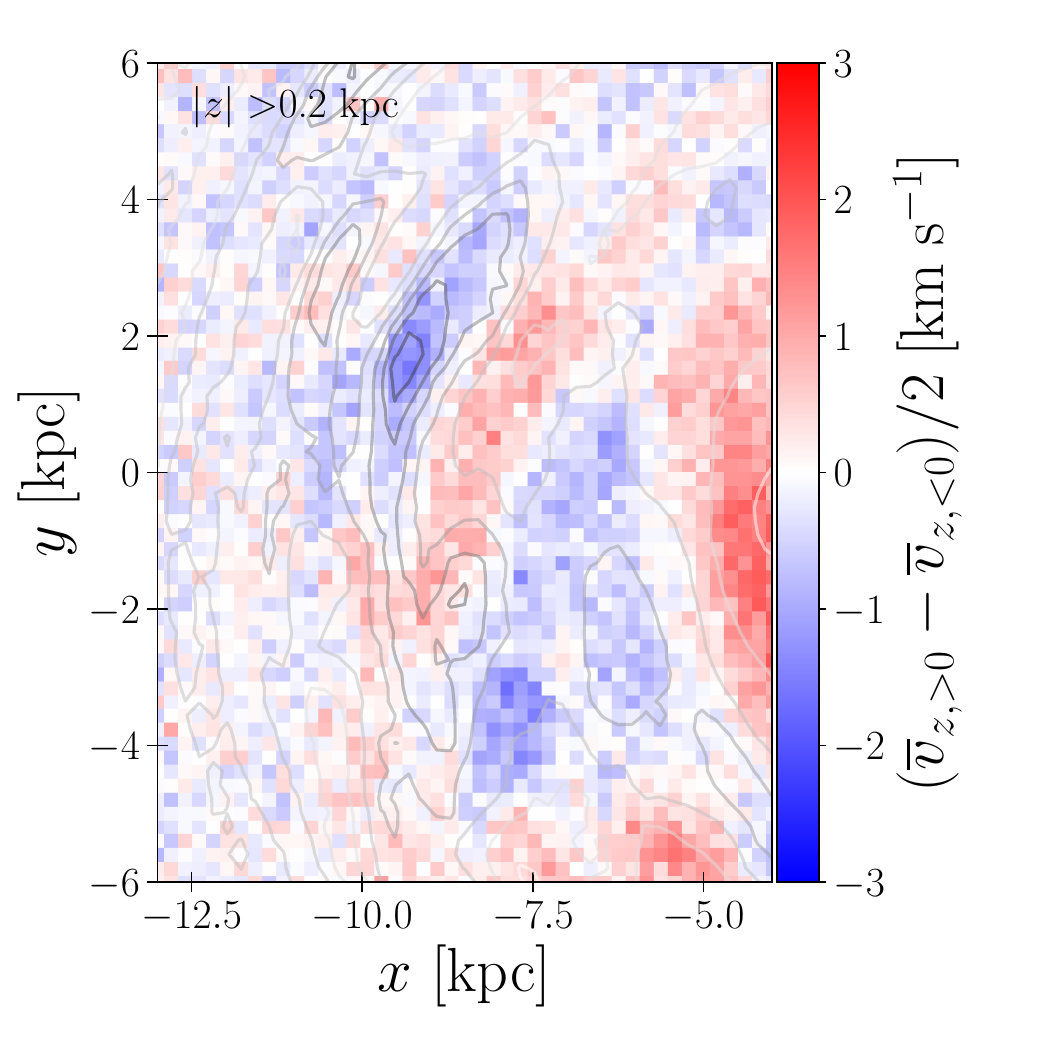}
		\includegraphics[width=0.49\columnwidth]{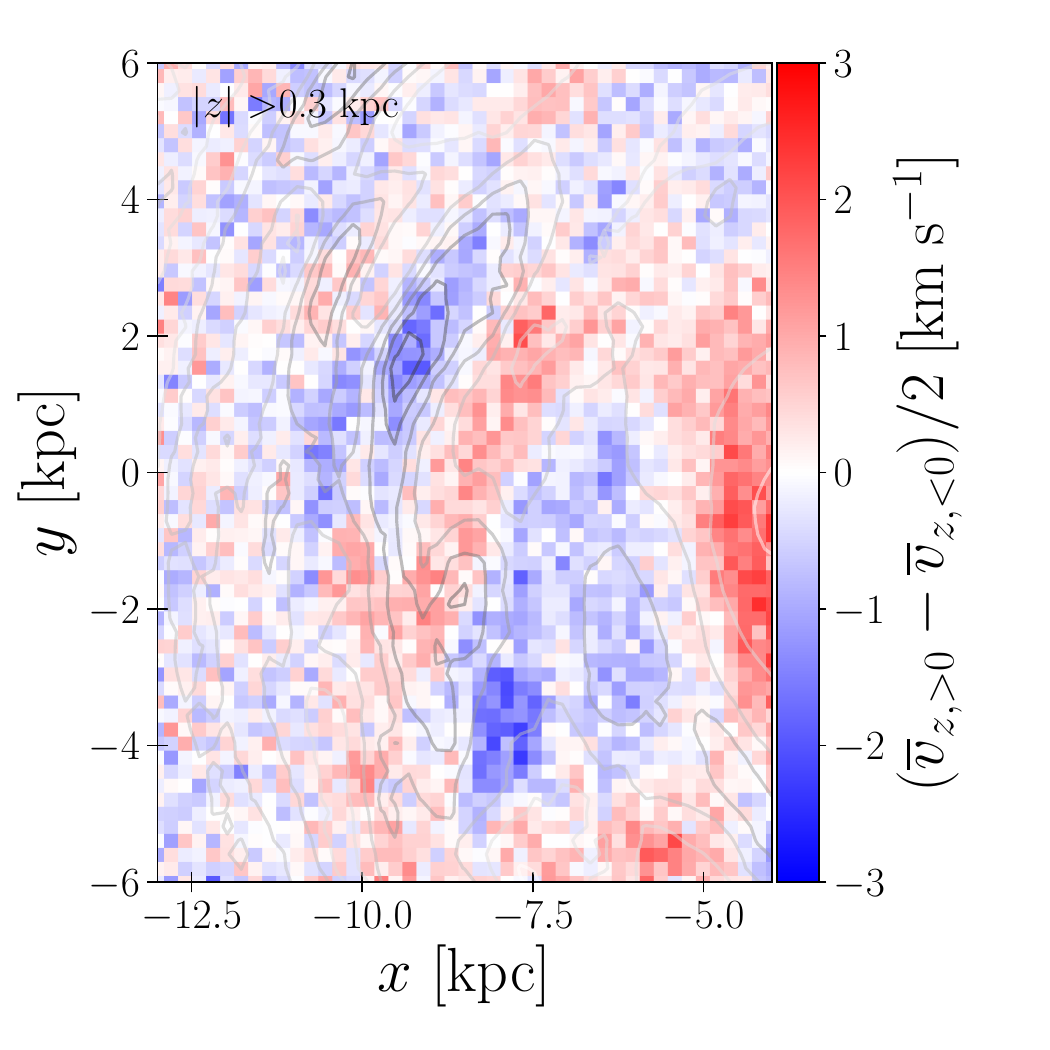}
		\includegraphics[width=0.49\columnwidth]{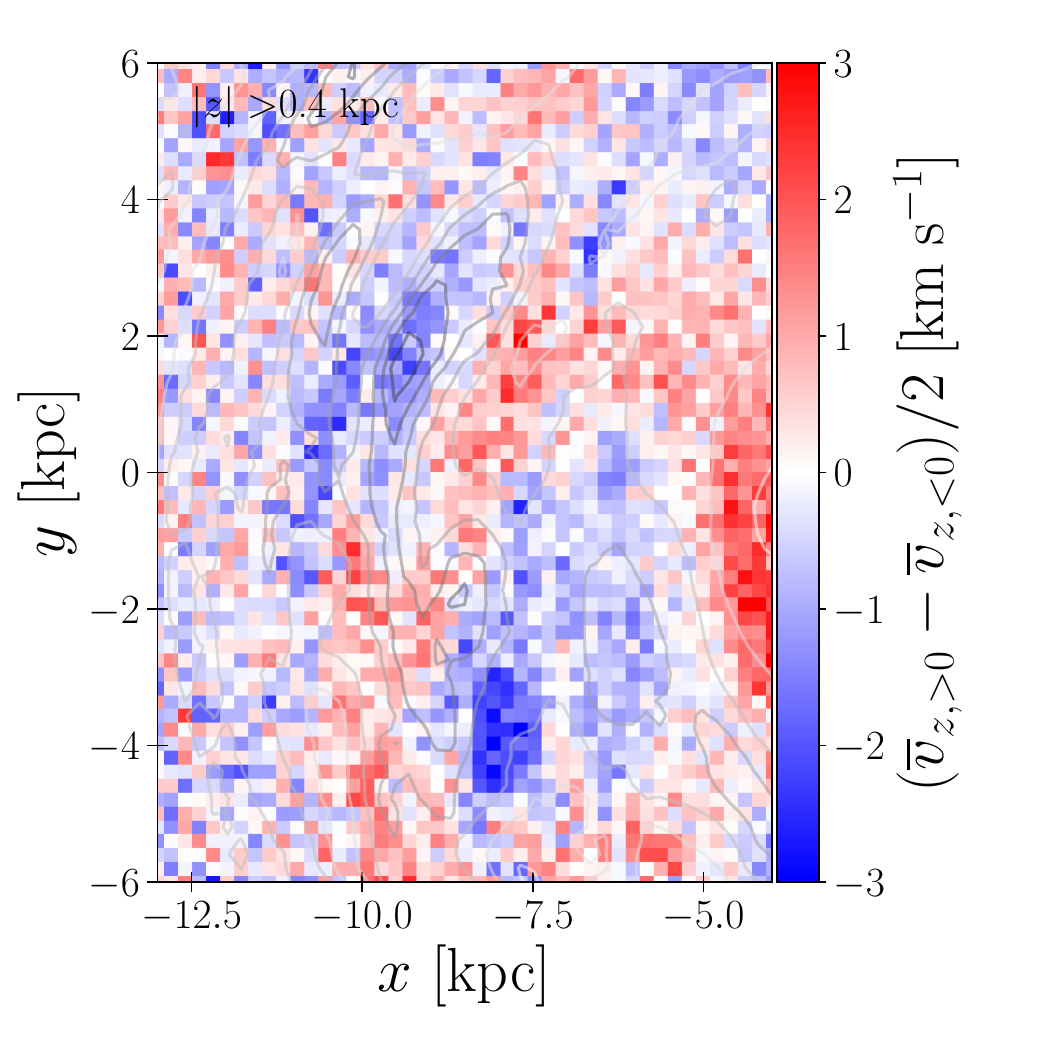}
		\caption{Face-on maps of the breathing velocity for the particles at $|z|>$0.1 (\textit{upper left}), 0.2 (\textit{upper right}), 0.3 (\textit{lower left}) and 0.4 (\textit{lower right}) kpc.}\label{fig:face_on_breath_zlim}
	\end{center}
\end{figure}
Fig.~\ref{fig:selection} shows the selection function for the \textit{Gaia} DR3 radial velocity sample \citep{2023A&A...669A..55C, 2023A&A...677A..37C} in the range of $60^{\circ} \le l \le 75^{\circ}$, corresponding to the line-of-sight direction to the Local arm. We plot the completeness as a function of $b$ for $G=11$, 13 and 15, with the $G-G_{RP}$ colour fixed at 1. We use \texttt{gaiaunlimited} Python package to make the plot.
The completeness decreases towards the mid-plane at $|b|\lesssim 5^{\circ}$. The Galactic latitude of $|b|=5^{\circ}$ corresponds to $|z|=$0.09, 0.17, 0.26, and 0.35 kpc at the heliocentric distances $d=$1, 2, 3 and 4 kpc, respectively.
Fig.~\ref{fig:face_on_breath_zlim} shows the face-on maps of the breathing velocity in the $N$-body model, as in the middle panel of Fig.~\ref{fig:face_on_MWa}, but here we use the particles only at $|z|>z_{\mathrm{lim}}$ ($=$ 0.1, 0.2, 0.3 and 0.4 kpc). This figure demonstrates how the selection bias affects the observed breathing velocity.
As the limiting height $z_{\mathrm{lim}}$ increases, the amplitude of the breathing velocity increases. The selection effect can explain the trend that the amplitude of the breathing velocity in the region distant from the Sun is larger than that in the solar neighbourhood as seen in Fig.~\ref{fig:face_on_Gaia}.

\bsp
\label{lastpage}
\end{document}